# Zitterbewegung at the level of quantum field theory*

WANG Zhi-Yong [†], XIONG Cai-Dong

*School of Optoelectronic Information, University of Electronic Science and Technology of China, Chengdu 610054, CHINA*

Traditionally, the zitterbewegung (ZB) of the Dirac electron has just been studied at the level of quantum mechanics. Seeing that the fact that an old interest in ZB has recently been rekindled by the investigations on spintronic, graphene, and superconducting systems, etc., in this paper we present a quantum-field-theory investigation on ZB and obtain the conclusion that, the ZB of an electron arises from the influence of virtual electron-positron pairs (or vacuum fluctuations) on the electron.

**Keywords:** Zittterbewegung; graphene; the Dirac electron

**PACC:** 3130J, 0365B

## 1. Introduction

An old interest in zitterbewegung (ZB) of the Dirac electron has recently been rekindled by the investigations on spintronics, graphene, and superconducting systems, etc. [1-6], where spintronics and graphene are red hot topics [7-9]. In particular, there are recently important progresses in improving the predictions for detecting ZB and relating them to Schrodinger cats in trapped ions [10, 11]. However, traditionally, ZB has been studied at the level of quantum mechanics [12-20], while a rigorous investigation on ZB at the field-quantized level is still absent. In this paper we will present a quantum-field-theory investigation for ZB, which has potential applications to spintronics, graphene, and superconducting systems, etc.

* Project supported by by the National Natural Science Foundation of China (Grant No. 60671030).
[†]Corresponding author.   E-mail: zywang@uestc.edu.cn

## 2. Zitterbewegung at the level of quantum mechanics

In the following, the natural units of measurement ($\hbar = c = 1$) is applied, repeated indices means summation according to the Einstein rule, and the four-dimensional (4D) space-time metric tensor is chosen as $g^{\mu\nu} = \text{diag}(1,-1,-1,-1)$, $\mu,\nu = 0,1,2,3$. At the level of quantum mechanics, there have been many investigations on the ZB of the Dirac electron [12-20]. Let $d^*$ denote the complex conjugate of $d$ (and so on), $\psi^\dagger$ denote the hermitian conjugate of the electron's wavefunction $\psi$ (and so on), as the general solution of the Dirac equation, the wavefunction $\psi$ can be written as

$$\psi(x) = \int \frac{d^3 p}{(2\pi)^{3/2}} \sqrt{\frac{m}{E}} \sum_s [c(p,s)u(p,s)\exp(-ip.x) + d^*(p,s)v(p,s)\exp(ip.x)], \qquad (1)$$

where $s = 1,2$ correspond to the spins of $\pm 1/2$, respectively, $p.x = p_\mu x^\mu$, $E = \sqrt{p^2 + m^2}$, $m$ is the mass of the Dirac electron, $u(p,s)$ and $v(p,s)$ are the 4×1 Dirac spinors in the momentum representation. Because all observables are the averages of operators rather than the operators themselves, one had better study the ZB of an electron via the electron's mean position $X = \int \psi^\dagger x \psi d^3 x$ and mean velocity $V = dX/dt$ (in the position space, the position operator $\hat{x} = x$). Using the Dirac equation $\hat{H}\psi = i\partial\psi/\partial t$ with $\hat{H} = \boldsymbol{\alpha} \cdot \hat{\boldsymbol{p}} + \beta m$ ($\hat{\boldsymbol{p}} = -i\nabla$, $\boldsymbol{\alpha}$ and $\beta$ are the Dirac matrices) and $\partial x/\partial t = 0$, one can prove that

$$V = dX/dt = \int \psi^\dagger (\partial x/\partial t + i[\hat{H}, x])\psi d^3 x = \int \psi^\dagger \boldsymbol{\alpha} \psi d^3 x. \qquad (2)$$

Substituting Eq. (1) into Eq. (2), one has $V = V_{\text{classic}} + V_{\text{zbw}}$, where

$$V_{\text{classic}} = \int d^3 p \sum_s [|c(p,s)|^2 + |d(p,s)|^2](\boldsymbol{p}/E) \qquad (3)$$

corresponds to the classic velocity of the Dirac electron, while [20]

$$V_{\text{zbw}} = \int d^3 p \{ \sum_{s,s'} [c^*(-p,s')d^*(p,s)\bar{u}(-p,s')\boldsymbol{\alpha} v(p,s)]\exp(i2Et) \\ - \sum_{s,s'} [d(-p,s')c(p,s)\bar{v}(-p,s')\boldsymbol{\alpha} u(p,s)]\exp(-i2Et) \} \qquad (4)$$

corresponds to the ZB velocity of the Dirac electron, where $c(p,s) \equiv c(E,\bm{p},s)$ and $c(-p,s) \equiv c(E,-\bm{p},s)$ (and so on), $\bar{u} = u^\dagger \beta$ (and so on). The ZB term is caused by the interference between the positive- and negative-energy components of the wavefunction $\psi$, and then it vanishes if there is only the positive- or negative-energy component.

## 3. Zitterbewegung at the level of quantum field theory

At the level of quantum field theory, the wavefunction $\psi$ should be reinterpreted as the field operator $\hat{\psi}$ in Fock space, where the operator property of $\hat{\psi}$ is carried by the creation and annihilation operators, then Eq. (1) becomes

$$\hat{\psi}(x) = \int \frac{d^3 p}{(2\pi)^{3/2}} \sqrt{\frac{m}{E}} \sum_s [\hat{c}(p,s) u(p,s) \exp(-ip.x) + \hat{d}^\dagger(p,s) v(p,s) \exp(ip.x)]. \qquad (5)$$

The expansion coefficients $\hat{c}^\dagger$ and $\hat{c}$ (or $\hat{d}^\dagger$ and $\hat{d}$) represent the electron's (or positron's) creation and annihilation operators, respectively. As mentioned before, at the level of quantum mechanics, the ZB can be studied via the electron's mean position $\bm{X} = \int \psi^\dagger \bm{x} \psi d^3 x$ and mean velocity $\bm{V} = d\bm{X}/dt = \int \psi^\dagger \bm{\alpha} \psi d^3 x$ (in the position space $\hat{\bm{x}} = \bm{x}$). However, in quantum field theory, the position vector $\bm{x}$ plays the role of a parameter rather than an operator. Nevertheless, in terms of the field operator $\hat{\psi}$ one can formally introduce an operator in Fock space as follows (its operator property is entirely carried by the creation and annihilation operators):

$$\hat{\bm{X}} \equiv \int \hat{\psi}^\dagger \bm{x} \hat{\psi} d^3 x = \hat{\bm{X}}_0 + \hat{\bm{X}}_1 + \hat{\bm{X}}_{z\perp} + \hat{\bm{X}}_{z\|}. \qquad (6)$$

In fact, let $q$ denote an electric charge, taking $\bm{x}$ as a displacement vector with respect to the origin of coordinates, one can regard $\hat{\bm{X}} = \int \hat{\psi}^\dagger q \bm{x} \hat{\psi} d^3 x$ as an electric dipole moment, it is a well-defined operator in quantum field theory. To calculate Eq. (6), let $\{\bm{e}_1, \bm{e}_2, \bm{e}_3\}$ denote an orthonormal basis with $\bm{e}_3 = \bm{e}_1 \times \bm{e}_2 = \bm{p}/|\bm{p}|$, its spinor representation forms another orthonormal basis $\{\boldsymbol{\eta}_+, \boldsymbol{\eta}_-, \boldsymbol{\eta}_\parallel\}$, where $\boldsymbol{\eta}_\pm = (\boldsymbol{e}_1 \pm i\boldsymbol{e}_2)/\sqrt{2}$ and $\boldsymbol{\eta}_\parallel = \boldsymbol{e}_3$ (see **Appendix A**), one can prove that

$$\hat{X}_0 = \sum_{p,s} (t\,\boldsymbol{p}/E)[\hat{c}^\dagger(p,s)\hat{c}(p,s) - \hat{d}^\dagger(p,s)\hat{d}(p,s)]. \tag{7}$$

$$\hat{X}_1 = \sum_{p,s} \{[(-\mathrm{i}\partial/\partial\boldsymbol{p})\hat{c}^+(p,s)]\hat{c}(p,s) + [(\mathrm{i}\partial/\partial\boldsymbol{p})\hat{d}(p,s)]\hat{d}^+(p,s)\}. \tag{8}$$

$$\hat{X}_{z\perp} = \sum_p \{\frac{-\mathrm{i}}{\sqrt{2E}} \boldsymbol{\eta}_+ [\hat{c}^\dagger(p,2)\hat{d}^\dagger(-p,1)\exp(\mathrm{i}2Et) + \hat{c}(-p,1)\hat{d}(p,2)\exp(-\mathrm{i}2Et)] + h.c.\}. \tag{9}$$

$$\hat{X}_{z\parallel} = \sum_p \frac{-\mathrm{i}m}{2E^2} \boldsymbol{\eta}_\parallel \{[\hat{c}^\dagger(p,1)\hat{d}^\dagger(-p,1) - \hat{c}^\dagger(p,2)\hat{d}^\dagger(-p,2)]\exp(\mathrm{i}2Et) + h.c.\}. \tag{10}$$

Here $h.c.$ denotes the hermitian conjugate of the preceding term, $\hat{c}(p,s) \equiv \hat{c}(E, \boldsymbol{p}, s)$ while $\hat{c}(-p, s) \equiv \hat{c}(E, -\boldsymbol{p}, s)$ (and so on). Obviously, $\hat{X}_0$ is related to the free motion of charges and describes the position of center-of-charge of the Dirac field; $\hat{X}_{z\perp}$ and $\hat{X}_{z\parallel}$ are perpendicular and parallel to the momentum vector $\boldsymbol{p}$, and arise from the contributions of the transverse and longitudinal ZB motions, respectively. If all particles have the same momentum (or velocity), one has $\hat{X}_1 = 0$, then $\hat{X}_1$ can be regarded as a correction for the position of center-of-charge of the Dirac field, which arise from a mutual effect between the fast and slow particles. Moreover, because $\hat{X}_1$, $\hat{X}_{z\perp}$, $\hat{X}_{z\parallel} \propto \hbar$ (note that we apply the natural units of measurement $\hbar = c = 1$), they are the quantum corrections for the position of center-of-charge of the Dirac field. Therefore, we call $\hat{X} \equiv \int \hat{\psi}^\dagger \boldsymbol{x} \hat{\psi} \mathrm{d}^3 x$ as the position operator of center-of-charge of the Dirac field, and then $\hat{V} = \mathrm{d}\hat{X}/\mathrm{d}t$ is the velocity operator of center-of-charge of the Dirac field. Obviously, $\hat{V} = \int \hat{\psi}^\dagger \boldsymbol{\alpha} \hat{\psi} \mathrm{d}^3 x = \int \hat{\boldsymbol{j}} \mathrm{d}^3 x$, where $\hat{\boldsymbol{j}} = \hat{\psi}^\dagger \boldsymbol{\alpha} \hat{\psi}$ is the 3D current density of the Dirac field, then $\hat{V} = \mathrm{d}\hat{X}/\mathrm{d}t$ also represents the 3D current operator of the Dirac field. One can prove that (the field $\hat{\psi}$ given by Eq. (5) is free, then $\mathrm{d}\hat{X}_1/\mathrm{d}t = 0$)

$$\hat{V} \equiv d\hat{X}/dt = \hat{V}_{classic} + \hat{Z}_\perp + \hat{Z}_\parallel, \tag{11}$$

where

$$\hat{V}_{classic} = d\hat{X}_0/dt = \sum_{p,s}(p/E)[\hat{c}^\dagger(p,s)\hat{c}(p,s) - \hat{d}^\dagger(p,s)\hat{d}(p,s)], \tag{12}$$

is the classic current, while $\hat{Z}_\perp = d\hat{X}_{z\perp}/dt$ and $\hat{Z}_\parallel = d\hat{X}_{z\parallel}/dt$ are the transverse and longitudinal ZB currents, respectively, they are,

$$\hat{Z}_\perp = \sum_p \{\sqrt{2}\eta_+[\hat{c}^\dagger(p,2)\hat{d}^\dagger(-p,1)\exp(i2Et) - \hat{c}(-p,1)\hat{d}(p,2)\exp(-i2Et)] + h.c.\}, \tag{13}$$

$$\hat{Z}_\parallel = \sum_p (m/E)\eta_\parallel \{[\hat{c}^\dagger(p,1)\hat{d}^\dagger(-p,1) - \hat{c}^\dagger(p,2)\hat{d}^\dagger(-p,2)]\exp(i2Et) + h.c.\}. \tag{14}$$

Eq. (12) shows that the classical current $\hat{V}_{classic}$ is formed by electrons or positrons with the momentum $p$. In contrary to which, in Eqs. (13) and (14), $\hat{c}^\dagger\hat{d}^\dagger$ and $\hat{c}\hat{d}$ are respectively the creation and annihilation operators of electron-positron pairs with vanishing total momentum (as viewed from any inertia frame of reference), then the ZB currents $\hat{Z}_\perp$ and $\hat{Z}_\parallel$ are related to the creation and annihilation of virtual electron-positron pairs. In fact, seeing that a hole in Dirac's hole theory can be interpreted as a positron, in terms of quantum field theory the traditional argument [19, 20] for the ZB of an electron (in a bound or free state) can be restated as follows: around an original electron, virtual electron-positron pairs are continuously created (and annihilated subsequently) in vacuum, the original electron can annihilate with the positron of a virtual pair, while the electron of the virtual pair which is left over now replaces the original electron, by such an exchange interaction the ZB occurs. Therefore, from the point of view of quantum field theory, the occurrence of the ZB for an electron arises from the influence of virtual electron-positron pairs (or vacuum fluctuations) on the electron.

Moreover, let $A_\mu$ be the 4D electromagnetic potential, $\hat{j}^\mu = (\hat{j}^0, \hat{\mathbf{j}})$ be the 4D

current-density vector, according to QED, in the electromagnetic interaction $\hat{j}^\mu A_\mu$ (let the unit charge $e=1$), the classical current $\hat{V}_{classic}$ can contribute to the Compton scattering, while the ZB currents $\hat{Z}_\perp$ and $\hat{Z}_\parallel$ can contribute to the Bhabha scattering. However, in the presence of the electromagnetic interaction, the vacuum is replaced with electromagnetic fields, and the electron-positron pairs in the Bhabha scattering are real rather than virtual ones.

## 4. Another interpretation for the position operator

As mentioned above, at the level of quantum field theory, $\hat{X} = \int \hat{\psi}^\dagger \mathbf{x} \hat{\psi} \mathrm{d}^3 \mathbf{x}$ can be interpreted as the position operator of center-of-charge of the Dirac field, and $\hat{V} = \mathrm{d}\hat{X}/\mathrm{d}t$ as the velocity operator of center-of-charge of the Dirac field, is also the Dirac current operator because of $\hat{V} = \int \hat{j} \mathrm{d}^3 \mathbf{x}$. Note that as the Fock-space operators, their operator property is entirely carried by the creation and annihilation operators, which avoids any problem with "position as an operator in quantum field theory". Now, let us provide another physical interpretation for $\hat{X} = \int \hat{\psi}^\dagger \mathbf{x} \hat{\psi} \mathrm{d}^3 \mathbf{x}$. As we know, the 4D momentum of fields acts as the conserved Noether charge related to the symmetry of the Poincaré group. Likewise, we will show that $\hat{X}^\mu = \int \hat{\psi}^\dagger x^\mu \hat{\psi} \mathrm{d}^3 \mathbf{x}$ can be regarded as a generalized (non-conserved) Noether charge associated with a local U(1) symmetry. For this let $A_\mu$ be the 4D electromagnetic potential, $\gamma^\mu$'s represent the Dirac matrices satisfying the algebra $\gamma^\mu \gamma^\nu + \gamma^\nu \gamma^\mu = 2g^{\mu\nu}$, $D_\mu = \partial_\mu + \mathrm{i}eA_\mu$ represent the covariant derivative (for convenience let the unit charge $e=1$), and $\bar{\psi} = \psi^\dagger \gamma^0$, according to QED, the Lagrangian density

$$\mathcal{L} = \hat{\bar{\psi}}(x)(\mathrm{i}\gamma^\mu D_\mu - m)\hat{\psi}(x) + (1/4)(\partial_\mu A_\nu - \partial_\nu A_\mu)(\partial^\mu A^\nu - \partial^\nu A^\mu). \tag{15}$$

is invariant under the local U(1) transformation as follows:

$$\hat{\psi}(x) \to \hat{\psi}'(x) = \exp[-i\theta(x)]\hat{\psi}(x), \quad A_\mu \to A'_\mu = A_\mu + \partial_\mu \theta(x), \qquad (16)$$

where $\theta(x)$ can be *arbitrary* scalar and continuously differentiable function. For our purpose, let $\theta(x) = \varepsilon_\mu x^\mu$, where $\varepsilon_\mu$'s ($\mu = 0,1,2,3$) are real constants. However, contrary to the usual local U(1) transformation, here we take $\varepsilon_\mu$ as the transformation parameter, while regard $x^\mu$ as the *generator* of the transformation, and call the transformation (16) with $\theta(x) = \varepsilon_\mu x^\mu$ a local pseudo-U(1)-transformation. Under the transformation (16), applying variation operations in $\delta \mathcal{L} = 0$ and applying the Euler-Lagrange equations of the fields, one can obtain an equation of continuity $D_\mu \hat{J}^{\mu\nu} = \hat{j}^\nu$, where

$$\hat{J}^{\mu\nu} = \hat{\bar{\psi}} \gamma^\mu x^\nu \hat{\psi}, \quad \hat{j}^\nu = \hat{\bar{\psi}} \gamma^\nu \hat{\psi}. \qquad (17)$$

In fact, applying Eq. (17) and $(i\gamma^\mu D_\mu - m)\hat{\psi} = 0$, $iD_\mu \hat{\bar{\psi}} \gamma^\mu + m\hat{\bar{\psi}} = 0$, one can examine the validity of $D_\mu \hat{J}^{\mu\nu} = \hat{j}^\nu$. By convention we take $\hat{X}^\nu \equiv \int \hat{J}^{0\nu} d^3 x$ as a charge, and call it a *generalized* Noether charge, it is

$$\hat{X}^\nu = \int \hat{\psi}^\dagger x^\nu \hat{\psi} d^3 x = (\hat{X}^0, \hat{\boldsymbol{X}}), \quad \text{with} \quad \hat{\boldsymbol{X}} = \int \hat{\psi}^\dagger \boldsymbol{x} \hat{\psi} d^3 x. \qquad (18)$$

As we know, under the global gauge transformation U(1), the free Lagrangian density $\hat{\bar{\psi}}(x)(i\gamma^\mu \partial_\mu - m)\hat{\psi}(x)$ is invariant, from which one can obtain the usual equation of continuity $\partial_\nu \hat{j}^\nu = 0$ and the conserved Noether charge $Q = \int \hat{j}^0 d^3 x$. Contrary to which, here the generalized Noether charge $\hat{X}^\mu = (\hat{X}^0, \hat{\boldsymbol{X}})$ is associated with a local symmetry and is no longer a conserved quantity: $d\hat{X}^\mu/dt \neq 0$, and its spatial component $\hat{\boldsymbol{X}} = \int \hat{\psi}^\dagger \boldsymbol{x} \hat{\psi} d^3 x$ is exactly the position operator of the center-of-charge of the Dirac field (defined by Eq. (6)). As $A_\mu = 0$, the equation of continuity becomes $\partial_\mu \hat{J}^{\mu\nu} = \hat{j}^\nu$, it can be examined via Eq. (17) and $(i\gamma^\mu \partial_\mu - m)\hat{\psi} = 0$, $i\partial_\mu \hat{\bar{\psi}} \gamma^\mu + m\hat{\bar{\psi}} = 0$.

## 5. Conclusions

Though the position vector $x$ in quantum field theory is just a parameter, one can still introduce a Fock-space operator $\hat{X} = \int \hat{\psi}^\dagger x \hat{\psi} \mathrm{d}^3 x$ (its operator property is entirely carried by the creation and annihilation operators), it plays the role of a generalized (non-conserved) Noether charge associated with a local U(1) symmetry of the Lagrangian density given by Eq. (15), and describes the position of center-of-charge of the Dirac field (it can also be regarded as the electric dipole moment with the electronic charge of $q=1$). Correspondingly, $\hat{V} = \mathrm{d}\hat{X}/\mathrm{d}t$ as the velocity of center-of-charge of the Dirac field, also represents the current vector of the Dirac field because of $\hat{V} = \int \hat{j} \mathrm{d}^3 x$. As a result, via $\hat{X} = \int \hat{\psi}^\dagger x \hat{\psi} \mathrm{d}^3 x$ and $\hat{V} = \mathrm{d}\hat{X}/\mathrm{d}t$ one can study the ZB of the Dirac electron at the level of quantum field theory, from which one can show that, from the point of view of quantum field theory, the ZB of an electron arises from the influence of virtual electron-positron pairs (or vacuum fluctuations) on the electron.

Acknowledgments: The first author (Wang Z Y) would like to thank professor Erasmo Recami for his useful discussions.

## Appendix A

In this paper, the orthonormal basis $\{e_1, e_2, e_3\}$ is expressed as:

$$\begin{cases} e_1 = e(p,1) = (\frac{p_1^2 p_3 + p_2^2 |p|}{|p|(p_1^2 + p_2^2)}, \frac{p_1 p_2 p_3 - p_1 p_2 |p|}{|p|(p_1^2 + p_2^2)}, -\frac{p_1}{|p|}) \\ e_2 = e(p,2) = (\frac{p_1 p_2 p_3 - p_1 p_2 |p|}{|p|(p_1^2 + p_2^2)}, \frac{p_2^2 p_3 + p_1^2 |p|}{|p|(p_1^2 + p_2^2)}, -\frac{p_2}{|p|}) \\ e_3 = e(p,3) = e(p,1) \times e(p,2) = \frac{p}{|p|} = \frac{1}{|p|}(p_1, p_2, p_3) \end{cases} \qquad (\mathrm{a}1)$$

As the spinor representation of $\{e_1, e_2, e_3\}$, another orthonormal basis $\{\eta_+, \eta_-, \eta_\parallel\}$ is defined by $\eta_\pm = (e_1 \pm ie_2)/\sqrt{2}$ and $\eta_\parallel = e_3$, that is

$$\begin{cases} \eta_+ = \eta_+(p) = (1/\sqrt{2}|p|)(\dfrac{p_1 p_3 - ip_2|p|}{p_1 - ip_2}, \dfrac{p_2 p_3 + ip_1|p|}{p_1 - ip_2}, -(p_1 + ip_2)) \\ \eta_- = \eta_-(p) = (1/\sqrt{2}|p|)(\dfrac{p_1 p_3 + ip_2|p|}{p_1 + ip_2}, \dfrac{p_2 p_3 - ip_1|p|}{p_1 + ip_2}, -(p_1 - ip_2)) \\ \eta_\parallel = \eta_\parallel(p) = e(p,3) = \dfrac{p}{|p|} = \dfrac{1}{|p|}(p_1, p_2, p_3) \end{cases} \quad (a2)$$

Obviously, the momentum $p$ is perpendicular to $\eta_\pm$ while parallel to $\eta_\parallel$. In fact, $\eta_\pm$ and $\eta_\parallel$ represent the three circular polarization vectors of the photon field (the spin-1 field), while $e_1$, $e_2$, and $e_3$ are the three linear polarization vectors. In terms of the matrix representation of the vectors $\eta_\pm$ and $\eta_\parallel$ ($\eta_-^*$ is the complex conjugate of $\eta_-$)

$$\eta_+ = \eta_-^* = \dfrac{1}{\sqrt{2}|p|} \begin{pmatrix} \dfrac{p_1 p_3 - ip_2|p|}{p_1 - ip_2} \\ \dfrac{p_2 p_3 + ip_1|p|}{p_1 - ip_2} \\ -(p_1 + ip_2) \end{pmatrix}, \quad \eta_\parallel = \dfrac{1}{|p|}\begin{pmatrix} p_1 \\ p_2 \\ p_3 \end{pmatrix}, \quad (a3)$$

one can prove that

$$\dfrac{\tau \cdot p}{|p|}\eta_i = \lambda_i \eta_i \quad (i = +, -, \parallel), \quad \lambda_\pm = \pm 1, \quad \lambda_\parallel = 0, \quad (a4)$$

where the spin-1 matrix vector $\tau = (\tau_1, \tau_2, \tau_3)$ has the components

$$\tau_1 = \begin{pmatrix} 0 & 0 & 0 \\ 0 & 0 & -i \\ 0 & i & 0 \end{pmatrix}, \quad \tau_2 = \begin{pmatrix} 0 & 0 & i \\ 0 & 0 & 0 \\ -i & 0 & 0 \end{pmatrix}, \quad \tau_3 = \begin{pmatrix} 0 & -i & 0 \\ i & 0 & 0 \\ 0 & 0 & 0 \end{pmatrix}. \quad (a5)$$

That is, $\eta_\pm$ and $\eta_\parallel$ represent the three eigenvectors of the spin-projection operator $\tau \cdot p/|p|$ of the spin-1 field, with the eigenvalues $\lambda_\pm = \pm 1$ and $\lambda_\parallel = 0$, respectively, which implies that $\eta_\parallel$ is the longitudinal polarization vector of the spin-1 field, while $\eta_+$ and $\eta_-$ are the right- and left-hand circular polarization vectors, respectively.